\documentclass[twocolumn]{aastex63}

\usepackage{xcolor}
\usepackage[normalem]{ulem}

\usepackage{amsmath}

\shorttitle{VLASS detected supernovae}
\shortauthors{Stroh et al.}

\begin{document}

\title{Luminous Late-time Radio Emission from Supernovae Detected by the Karl G.\ Jansky Very Large Array Sky Survey (VLASS)}

\correspondingauthor{M.C.\ Stroh}
\email{michael.stroh@northwestern.edu}

\author[0000-0002-3019-4577]{Michael C. Stroh}
\affiliation{Center for Interdisciplinary Exploration and Research in Astrophysics (CIERA) and Department of Physics and Astronomy, Northwestern University, Evanston, IL 60201, USA
}

\author[0000-0003-0794-5982]{Giacomo Terreran}
\affiliation{Center for Interdisciplinary Exploration and Research in Astrophysics (CIERA) and Department of Physics and Astronomy, Northwestern University, Evanston, IL 60201, USA
}
\affiliation{Las Cumbres Observatory, 6740 Cortona Drive, Suite 102, Goleta, CA 93117-5575, USA}

\author[0000-0001-5126-6237]{Deanne L. Coppejans}
\affiliation{Center for Interdisciplinary Exploration and Research in Astrophysics (CIERA) and Department of Physics and Astronomy, Northwestern University, Evanston, IL 60201, USA
}
\affiliation{Department of Physics, University of Warwick, Gibbet Hill Road, Coventry CV4 7AL, UK}

\author[0000-0002-7735-5796]{Joe S. Bright}
\affiliation{Center for Interdisciplinary Exploration and Research in Astrophysics (CIERA) and Department of Physics and Astronomy, Northwestern University, Evanston, IL 60201, USA
}
\affiliation{Department of Astronomy, University of California, Berkeley, CA 94720, USA}

\author[0000-0003-4768-7586]{Raffaella Margutti}
\affiliation{Center for Interdisciplinary Exploration and Research in Astrophysics (CIERA) and Department of Physics and Astronomy, Northwestern University, Evanston, IL 60201, USA
}
\affiliation{Department of Astronomy, University of California, Berkeley, CA 94720, USA}

\author[0000-0002-0592-4152]{Michael F. Bietenholz}
\affiliation{Department of Physics and Astronomy, York University, Toronto, M3J 1P3, Ontario, Canada
}
\affiliation{Hartebeesthoek Radio Observatory, P.O.\ Box 443, Krugersdorp, 1740, South Africa
}

\author[0000-0002-3137-4633]{Fabio De Colle}
\affiliation{Instituto de Ciencias Nucleares, Universidad Nacional Aut\'onoma de M\'exico, A.\ P.\ 70-543 04510 D.F.\ Mexico}

\author[0000-0003-4587-2366]{Lindsay DeMarchi}
\affiliation{Center for Interdisciplinary Exploration and Research in Astrophysics (CIERA) and Department of Physics and Astronomy, Northwestern University, Evanston, IL 60201, USA
}

\author[0000-0002-5565-4824]{Rodolfo Barniol Duran}
\affiliation{Department of Physics and Astronomy, California State University, Sacramento, 6000 J Street, Sacramento, CA 95819, USA
}

\author[0000-0002-0763-3885]{Danny Milisavljevic}
\affiliation{Department of Physics and Astronomy, Purdue University, 525 Northwestern Avenue, West Lafayette, IN 47907, USA
}

\author[0000-0002-5358-5642]{Kohta Murase}
\affiliation{Department of Physics, The Pennsylvania State University, University Park, PA 16802, USA}
\affiliation{Department of Astronomy \& Astrophysics, The Pennsylvania State University, University Park, PA 16802, USA}
\affiliation{Center for Mulitmessenger Astrophysics, Institute for Gravitation and the Cosmos, The Pennsylvania State University, University Park, PA 16802, USA}
\affiliation{Center for Gravitational Physics, Yukawa Institute for Theoretical Physics, Kyoto University, Sakyo-ku, Kyoto 606-8502, Japan}

\author[0000-0001-8340-3486]{Kerry Paterson}
\affiliation{Center for Interdisciplinary Exploration and Research in Astrophysics (CIERA) and Department of Physics and Astronomy, Northwestern University, Evanston, IL 60201, USA
}

\author[0000-0001-7315-1596]{Wendy L. Williams}
\affiliation{Leiden Observatory, Leiden University, PO Box 9513, NL-2300 RA Leiden, the Netherlands}

\begin{abstract}
We present a population of 19 radio-luminous supernovae (SNe) with emission reaching $L_{\nu}{\sim}10^{26}-10^{29}\,\rm{erg\,s^{-1}Hz^{-1}}$ in the first epoch of the Very Large Array Sky Survey (VLASS) at $2-4$\,GHz. Our sample includes one long Gamma-Ray Burst, SN\,2017iuk/GRB171205A, and 18 core-collapse SNe detected at $\approx (1-60)$\,years after explosion. No thermonuclear explosion shows evidence for bright radio emission, and hydrogen-poor progenitors dominate the sub-sample of core-collapse events with spectroscopic classification at the time of explosion (79\%).  We interpret these findings into the context of the expected radio emission from the forward shock interaction with the circumstellar medium (CSM). We conclude that these observations require a departure from the single wind-like density profile (i.e., $\rho_{\rm{CSM}}\propto r^{-2}$) that is expected around massive stars and/or a departure from a spherical Newtonian shock. 
Viable alternatives include the shock interaction with a detached, dense shell of CSM formed by a large effective progenitor mass-loss rate $\dot M \sim (10^{-4}-10^{-1})$\,M$_{\odot}$\,yr$^{-1}$ (for an assumed wind velocity of $1000\,\rm{km\,s^{-1}}$); emission from an off-axis relativistic jet entering our line of sight; or the emergence of emission from a newly-born pulsar-wind nebula. The relativistic SN\,2012ap that is detected 5.7 and 8.5\,years after explosion with $L_{\nu}{\sim}10^{28}$\,erg\,s$^{-1}$\,Hz$^{-1}$ might constitute the first detections of an off-axis jet+cocoon system in a massive star.
However, none of the VLASS-SNe with archival data points are consistent with our model off-axis jet light curves.
Future multi-wavelength observations will distinguish among these scenarios.
Our VLASS source catalogs, which were used to perform the VLASS cross matching, are publicly available at\dataset[10.5281/zenodo.4895112]{https://doi.org/10.5281/zenodo.4895112}.
\end{abstract}


\section{Introduction}
\label{sec:intro}
Radio observations of stellar explosions in the years-to-decades after stellar demise constitute a probe of the physical properties of the fastest ejecta in the explosion (i.e. their velocity and energy), and of the environment at large distances of $r\ge 10^{17}\,\rm{cm}$ (i.e., the density of the circumstellar medium, CSM), e.g., \cite{Chevalier2017}. There are three main sources of bright non-thermal synchrotron radio emission in SNe at $t\ge 1$ yr: (i) the deceleration of the forward shock in a dense environment \citep[e.g.,][]{1998ApJ...499..810C,2006ApJ...651..381C}; (ii) emission from an off-axis relativistic jet entering our line of sight \citep[e.g.,][]{Granot02}; (iii) emergence of emission from a newly-formed pulsar-wind nebula \citep[PWN;][]{Slane2017}. 
Late-time radio observations of cosmic explosions can thus reveal a complex mass-loss history of the stellar progenitors in the years leading up to core-collapse; they can reveal jet-driven explosions similar to long Gamma-Ray Bursts (GRBs) that launched a jet that was misaligned with our line of sight; or they can reveal the energetics and properties of the compact object remnant. 
However, most SNe are not observed at radio wavelengths at very late times. 
For example, in the sample of 294 SNe observed at $\sim5-8$\,GHz compiled by \citet{Bietenholz_2021}, only 87 were observed at more than 1000 days post explosion and of these only 28 were detected\footnote{This sample is one of the largest compilations of radio observations of SNe. 
We note that it is not a complete sample (see \citealt{Bietenholz_2021}), so these numbers are likely underestimates. 
One important selection effect is that the peak frequency of the radio emission from SNe declines with time, so lower frequency observations would yield a higher detection rate.}.
As a result, the late-time radio emission from SNe constitutes a poorly explored region of the phase space (Figure 2 in \citealt{Bietenholz_2021}). 

Here we present a sample of 19 radio-luminous SNe detected in the first epoch of the Very Large Array Sky Survey \citep[VLASS,][]{2020PASP..132c5001L} carried out by the Karl G.\ Jansky Very Large Array (VLA).
VLASS is a successor and complementary survey to the National Radio Astronomy Observatory (NRAO) VLA Sky Survey \citep[NVSS,][]{1998AJ....115.1693C} and Faint Images of the Radio Sky at Twenty centimeters \citep[FIRST,][]{1995ApJ...450..559B} surveys. The survey is conducted at $2$ to $4$\,GHz and is split into three distinct epochs, each scanning the full survey region (declination $\delta {>}{-}40^{\circ}$) with an ${\approx}\,32$\,month observing cadence reaching an RMS noise of $120\mu$Jy/beam per epoch.
Exploring `Hidden Explosions and Transient Events' is a key VLASS science theme, and \citet{2020ATel14020....1H} demonstrated the synergy between VLASS and newly discovered transients with the detection of the type II SN\,2019xhb in the 2nd VLASS observing epoch 202 days after discovery.  Importantly, by scanning the northern sky, VLASS offers the opportunity to perform a systematic and unbiased survey of the late-time radio emission from the tens of thousands of previously reported SNe. With the exception of SN\,2017iuk/GRB171205A, the sample of radio luminous SNe presented in this paper was imaged at an epoch corresponding to $\approx (1-60)$\,years post explosion.

This paper is organized as follows. In Section \ref{sec:sample}, we describe our methodology for identifying optically detected SNe in the VLASS sample, how we filtered out potential spurious detections, and present our list of VLASS detected SNe. Section \ref{sec: powering} discusses physical processes that could produce the bright radio emission associated with the SNe. Finally, in Section \ref{sec:conclusions} we summarize the conclusions.


\section{A Sample of SNe with luminous late-time radio emission}
\label{sec:sample}
We created a catalog consisting of all publicly-announced, optical SNe by combining the Bright Supernovae\footnote{\href{http://rochesterastronomy.org/snimages}{http://rochesterastronomy.org/snimages}}, Open SNe\footnote{\href{https://sne.space}{https://sne.space}}, and Transient Name Server\footnote{\href{https://www.wis-tns.org}{https://www.wis-tns.org}} catalogs.
We included all optical SNe detected prior to 2020-01-01 leading to an initial sample of $\approx$70,000 unique SNe. 
Two independently generated VLASS detection catalogs were produced using the ${\approx}35,000$ VLASS epoch 1 quick look images provided by NRAO.

\subsection{PyBDSF detections}
\label{sec:sample_pybdsf}
The complete VLASS epoch 1 was processed using the Python Blob Detector and Source Finder version 1.9.1 \citep[\texttt{PyBDSF},][]{2015ascl.soft02007M}.
We created a local catalog consisting of all detections using a source detection threshold of 5$\sigma$ (\texttt{thresh\_pix=5.0}), threshold for islands of 3$\sigma$ (\texttt{thresh\_isl=3.0}), and fixing source components to be Gaussian with major axis, minor axis, and position angle equal to the synthesised beam shape from the respective VLASS observations (\texttt{fix\_to\_beam=True}). The values of \texttt{thresh\_pix} and \texttt{thresh\_isl} are used to calculate pixel islands of significant emission. The \texttt{thresh\_pix} parameter is used to identify significant pixels (where the pixel value is greater than mean + $\texttt{thresh\_pix}\times\sigma$, where $\sigma$ is the rms noise of the image) used for fitting, with the mean map calculated using a box of pixel size and pixel step size either calculated within \texttt{PyBDSF} or set manually by the user. The fitting region is then extended based on the value of \texttt{thresh\_isl} such that all pixels greater than mean + $\texttt{thresh\_isl}\times\sigma$ which are adjacent (including diagonally) to a significant pixel identified using \texttt{thresh\_pix} are included in the fitting region or `island' (if this process causes islands to overlap they are combined). Multiple Gaussians are then fit to the island in order to best describe the source (the collection of Gaussians within an island) and we fixed the Gaussians to the shape of the synthesized beam.

This VLASS-PyBDSF catalog resulted in 3~752~214 sources, similar to the 3~381~277 VLASS sources cataloged by the Canadian Initiative for Radio Astronomy Data Analysis \citep[CIRADA,][]{2020RNAAS...4..175G}, and more than the 2~232~726 sources in the VLASS Quicklook Catalog \citep{2021ApJ...914...42B}.
Our VLASS-PyBDSF catalog is publicly available at\dataset[10.5281/zenodo.4895112]{https://doi.org/10.5281/zenodo.4895112}.

\begin{figure*}[t!]
\centering
  \includegraphics[width=\textwidth]{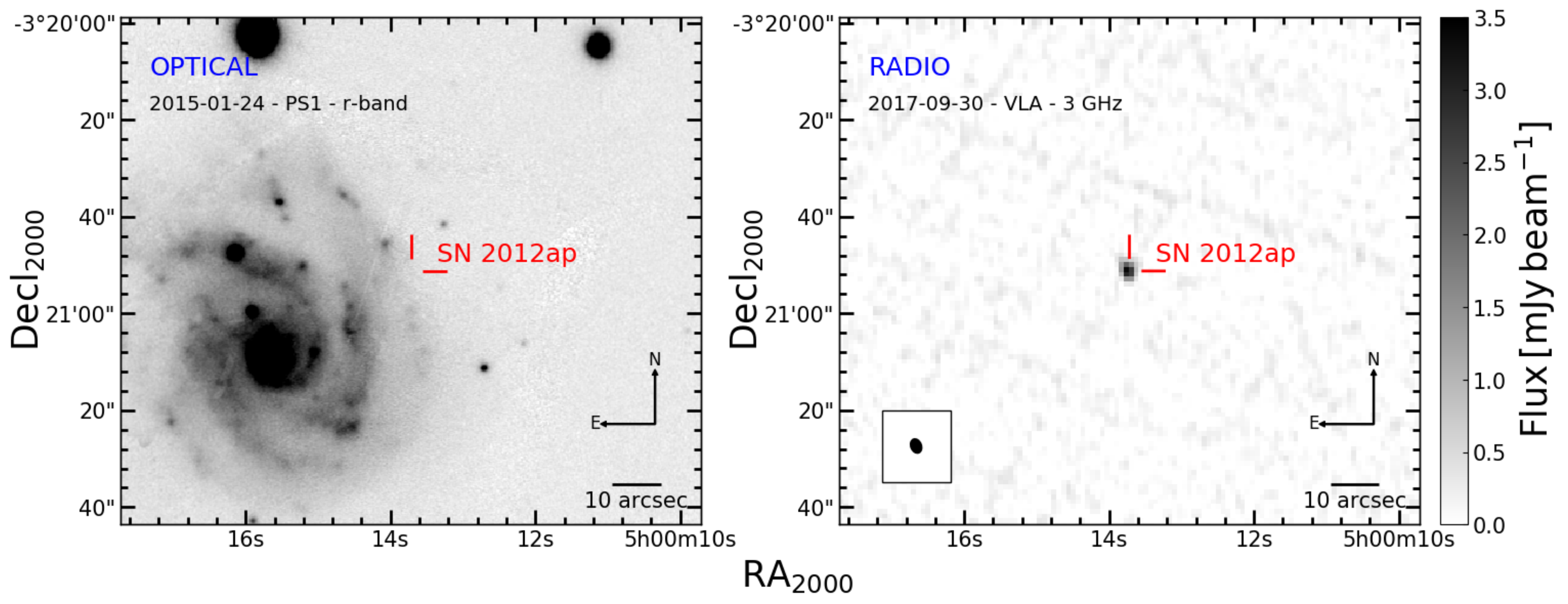}
  \caption{ 
  Field of SN\,2012ap in the Pan-STARRS1 r-band (left) and at 3\,GHz with VLASS (right). 
  The red lines indicate the optical SN position, and each image notes the date of each observation.
  The optical image clearly shows the host galaxy of SN\,2012ap, but no emission from the host is found in the optical image near the optical SN position.
  No radio emission is found in the field near SN\,2012ap except near the optical SN position.
  The VLASS image was taken 2065 days since explosion \citep{2014ApJ...782L...5M}.
  }
\label{fig:SN2012ap_example}
\end{figure*}

\subsection{Source extractor detections}
\label{sec:sample_sextractor}
We created an initial list of possible VLASS source detections using Source Extractor version 2.25.0 \citep[\texttt{SExtractor},][]{1996A&AS..117..393B}.
\texttt{SExtractor} is not optimized for radio imaging analysis, but through trial and error with fields where transients were known to exist, we settled on a requirement of at least 5 contiguous pixels (\texttt{detect\_minarea}) to be above the low \texttt{detect\_threshold=5}$\sigma$.
The VLASS-SExtractor catalog contained 9~652~665 possible source detections.
Most of these detections are likely spurious, as evidenced by the number of sources detected by the previous section.
We note that \texttt{SExtractor} runs at least an order of magnitude faster through VLASS quick look images than \texttt{PyBDSF}, so it may be preferred for studies where a large number of possible bogus detections are acceptable.
Our VLASS-SExtractor catalog is publicly available at\dataset[10.5281/zenodo.4895112]{https://doi.org/10.5281/zenodo.4895112}.

\subsection{Identifying supernovae for cross matching}
We chose low detection thresholds in order to minimize the chance that we may miss potential SNe cross matches on our initial pass.
In order to reduce the number of spurious matches, we required that the potential SNe cross matches must be detected by both VLASS-SExtractor and VLASS-PyBDSF catalogs.
We cross-matched the locations of the optical SNe with the VLASS-PyBDSF and VLASS-SExtractor catalogs using a $5''$ angular separation.
This separation is ${\approx}2$ times the average VLASS beam size, and helps account for a lack of positional precision in SN discovery reports. 
Using our initial list of ${\approx}70,000$ SNe, ${\approx}1600$, and ${\approx}1400$ SNe have potential PyBDSF, and SExtractor cross-matches, respectively. 
By requiring that a source must have cross-matches in both the VLASS-PyBDSF and VLASS-SExtractor catalogs, we have ${\approx}1300$ potential cross-matches in VLASS.
We further required that the VLASS observation must have occurred \emph{after} the SN discovery date.

The possible VLASS-SNe were visually inspected to ensure that the VLASS-SNe detections are real and are not due to radio imaging artifacts (see for example Figure \ref{fig:SN2012ap_example}). 
We also rejected SNe when the location of the radio source broadly overlapped with that of the galactic nucleus. 
After visual inspection, only ${\approx}100$ potential VLASS-SNe detections remained.

\subsection{Multiwavelength cross matching} \label{sec:multiwavelength_filtering}
In order to filter out known radio sources, we rejected associations that had counterparts in the NVSS or FIRST catalogs prior to their explosion date.
We removed VLASS-SNe near active galactic nuclei (AGN) identified by \citet{2018ApJS..234...23A}, who cataloged probable AGN in the \textit{Wide-field Infrared Survey Explorer} \citep[WISE,][]{2010AJ....140.1868W} AllWISE data release \citep{2014yCat.2328....0C}. 

\citet{2013ApJS..206...12D} examined the chance of probability for spurious associations between a sample of NVSS detected blazars and AllWISE.
They calculated the number of additional cross-matches between their NVSS blazar catalog and AllWISE $\Delta N_t$ as the cross-matching radius increases.
Similarly, the number of additional spurious cross-matches per increasing cross-matching radius, $\Delta N_r$, was calculated by adding a random offset to the NVSS blazar positions.
\citet{2013ApJS..206...12D} found that $\Delta N_r > \Delta N_t$ for cross-matching radii above $3.3''$, thus a cross-matching radius of $3.3''$ can be considered a cross-matching between point-like VLA sources and AllWISE.
For the infrared AGN cross matching, we adopted the $3.3''$ angular search radius suggested by \citet{2013ApJS..206...12D}.

Possibly misidentified AGN were also removed by cross matching our VLASS-SNe candidates against the \textit{Chandra} Source Catalog v2.0 \citep[CSC 2.0,][]{2020AAS...23515405E}, the most recent \textit{XMM-Newton} X-ray source data release \citep[4XMM-DR10,][]{2020A&A...641A.136W}, and the 2nd \textit{Swift}-XRT point source catalog \citep[2SXPS,][]{2020ApJS..247...54E}.
The error in the X-ray position is generally greater than the astrometric uncertainties in the VLASS positions \citep[see the Appendix of][]{2021ApJ...914...42B}, thus for X-ray catalog cross-matching, we removed sources within the 1$\sigma$ X-ray error region in the respective X-ray catalog.
We ensured that no VLASS-SNe candidates were rejected by targeted SNe follow-up observations. 


We further inspected the VLASS-SNe candidates within the 424 square degrees covered so far by the LOFAR Two-metre Sky Survey \citep[LoTSS Data Release 1,][]{2019A&A...622A...1S}, to ensure that none were classified as AGN based on any of their $150$\,MHz radio morphology or luminosity, or based on their multi-wavelength cross-identifications \citep{2019A&A...622A...2W}. 
While only two of the candidates lie within this area, both were detected as star-forming galaxies. 
Further releases of LoTSS over the Northern sky will enable further such comparisons.

This multiwavelength filtering procedure leads to a sample of 19 core-collapse SNe with associated VLASS emission.
The final VLASS-SNe sample is listed in Table \ref{tab:final_detection_list}.
All VLASS-SNe also have counterparts in the CIRADA VLASS catalogue, and SN\,2017hcb is the only source without a cross-match in the VLASS Quicklook Catalog.
Interestingly, we note that no thermonuclear explosion (i.e.\ Ia-like) passed the criteria above, in spite of it largely dominating the initial optical SN sample. 
The lack of type Ia SNe in the sample is consistent with the lack of radio emission associated with type Ia \citep[e.g.,][]{Chomiuk_2016}, but the lack chance coincidence matches may be evidence of the strength of our multiwavelength filtering described above.
Of the 14 VLASS detected SNe with early-time spectroscopic classification, 13 (93\%) and 12 (86\%) are detected at ${>}10^2$ and ${>}10^3$\,days post explosion, respectively.
SNe with hydrogen poor progenitors at the time of explosion make-up the majority 11 (79\%) of the sources with early-time spectroscopic classification.
Remarkably, we find that SN\,2012ap, one of the only two known SNe with relativistic ejecta without a GRB \citep{Margutti_2014,2015ApJ...805..187C,Milisavljevic_2015}, shows evidence for bright radio emission years after explosion, and is a member of the sample.
The very nearby SN\,2017iuk/GRB171205A, at redshift $z{=}0.0368$ \citep{2017TNSCR1387....1D}, was detected in the first VLASS epoch less than 60 days following the \textit{Swift} trigger.

\subsection{Final flux densities}
\label{sec:sample_final_list}
In addition to a complete and consistent processing of VLASS epoch 1 (as described in Section \ref{sec:sample_pybdsf}) we performed an optimized manual analysis on each of the fields containing the sources listed in Table \ref{tab:final_detection_list}. 
The default significance parameters used for this analysis were \texttt{thresh\_isl=5.0} and \texttt{thresh\_pix=5.0}, with an adaptive region used to calculate the RMS and mean maps (\texttt{adaptive\_rms\_box=True}), and components fixed to be the same shape as the synthesised beam (\texttt{fix\_to\_beam=True}). The island and pixel threshold values were adjusted on a per-field basis depending on e.g.\ bright source artifacts or extended host structure; however, we require $\texttt{thresh\_pix}\geq\texttt{thresh\_isl}$. Adaptive RMS calculation ensures that region size near bright sources is reduced compared to regions devoid of bright emission, properly accounting for elevated RMS levels resulting from imaging artifacts (an issue in a number of the VLASS fields). For fields with particularly strong artifacts we set the RMS box size and step size manually such that the noise map captured the variation caused by the artifacts. In the cases where the SN emission formed part of an extended emission complex from the host galaxy we set \texttt{fix\_to\_beam=False} in order to better model the emitting region. 
We ran \texttt{PyBDSF} in interactive mode (\texttt{interactive=True}) and manually inspected the result of the island and source detection, adjusting our significance threshold and the size of the region used to calculate the RMS noise to improve the source fitting. 

\begin{figure*}[t!]
\centering
  \includegraphics[width=\textwidth]{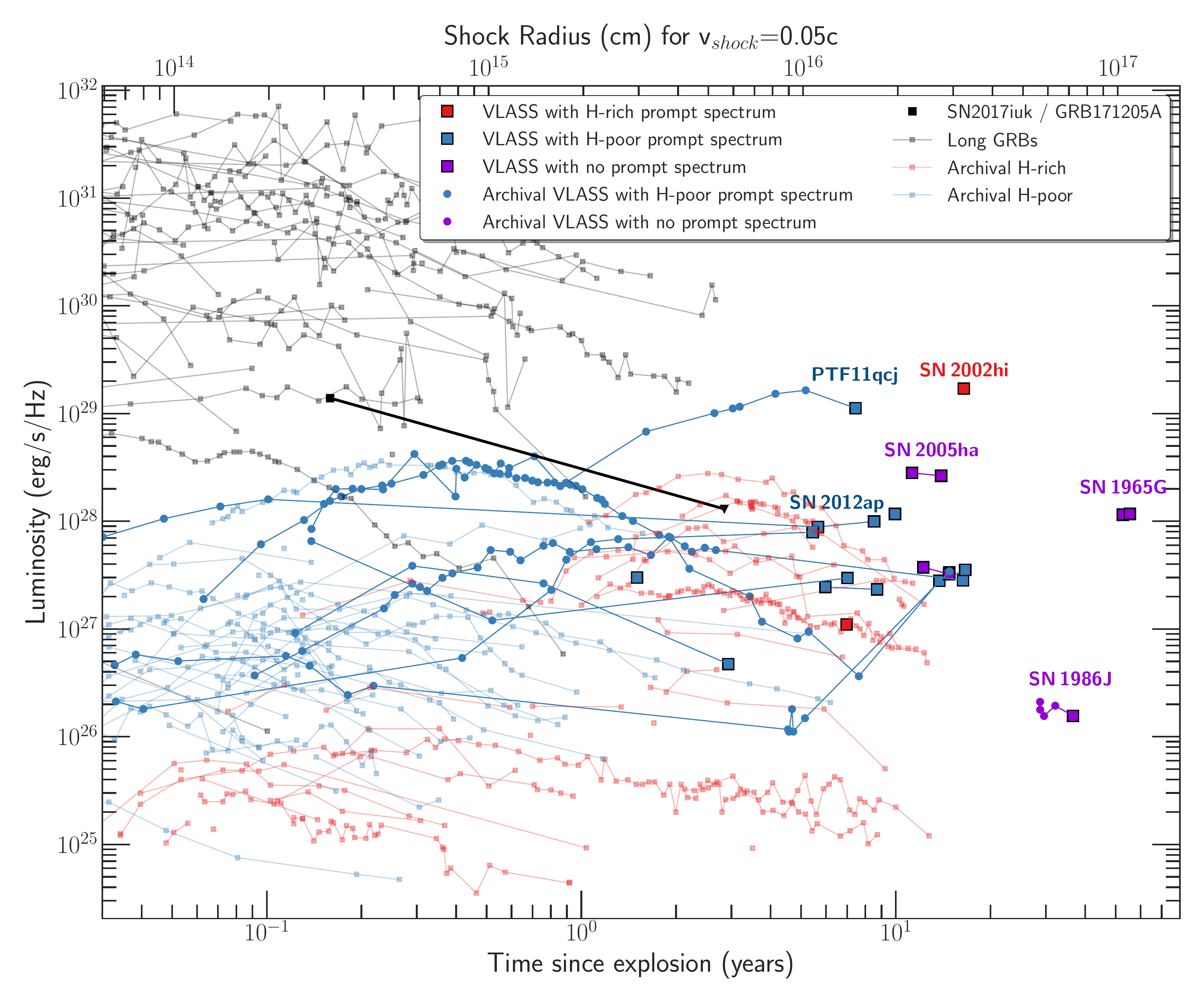}
  \caption{ 
  VLASS-SNe detections in the context of H-rich SNe (red), H-poor SNe (blue) SNe and long GRBs (gray). 
  A number of the VLASS observations were taken at a later stage than SNe are typically observed and detected at radio wavelengths, and show brighter emission than would be expected at this epoch. 
  Notably, the H-poor VLASS-SNe were observed years after the radio emission (at $\gtrsim$1 GHz) from this class typically fades.
  Archival radio light curves for VLASS detected SNe are included: SN\,1986J \citep[$3-5$\,GHz;][]{Bietenholz17b}, 
  SN\,2003bg \citep[4.86\,GHz;][]{2006ApJ...651.1005S}, 
  SN\,2004C (4.9\,GHz; DeMarchi in prep.), 
  SN\,2004dk \citep[$3-5$\,GHz;][]{2012ApJ...752...17W,Balasubramanian_2021}, 
  PTF11qcj \citep[$3-4$\,GHz;][]{Palliyaguru_2019,2014ApJ...782...42C}, 
  SN\,2012ap \citep[$3$\,GHz extrapolation based on radio SED modeling;][]{2015ApJ...805..187C},
  SN\,2012au \citep[$3-4$\,GHz;][Terreran et al.\ in prep.]{Kamble_2014}, 
  SN\,2014C \citep[7.1\,GHz;][]{Margutti17}, and
  SN\,2016coi \citep[3\,GHz;][]{Terreran_2019}.
  The archival radio observations of SNe are from \citet{Bietenholz_2021}, and the archival long GRBs are from \citet{Chandra_2012}. 
  Most archival historical light curves are at 8.6\,GHz, as the 3 GHz light curves are not well sampled.
  From the VLASS detected sample, SDSS-II SN\,8524 is not included since it has neither a known host galaxy nor redshift, thus a luminosity cannot be calculated. 
  The H-rich and H-poor designations are inferred from the spectrum near the time of explosion.
  The upper x-axis provides a reference distance scale for a fiducial normal SN shock velocity of $0.05$c with no deceleration.
  This figure highlights the presence of two groups of H-rich SNe in the radio phase space, with IIn SNe belonging to the group with luminous radio emission years after explosion \citep[see e.g.\ ][]{Bietenholz_2021}.
  }
\label{fig:lum_function}
\end{figure*}

For fields with extreme imaging artifacts around bright sources, we manually set the RMS box size and disabled adaptive region sizing. 
In a handful of cases, emission from the transient was part of a larger emission complex (radio emission from the host galaxies) and the emission island was better described using Gaussians with unconstrained shapes. 
We list the results of our fitting in Table \ref{tab:final_detection_list}, and note any deviations from the default parameters. 
Additionally we analyzed each of the target fields and list the flux densities and position of the sources in the first half of the second VLASS epoch (i.e. epoch 2.1). 
In the cases of SN\,2017iuk/GRB171205A and SDSS-II SN\,8524, the source is no longer detected in the second epoch, so instead we list a $3\sigma$ upper limit.
We compare the luminosities and timescales of the VLASS detected SNe to historical radio light curves in Figure \ref{fig:lum_function}.

\section{Powering luminous late-time radio emission }
\label{sec: powering}

Winds from massive stars enhance and shape the density of their immediate surroundings \citep[e.g.,][]{smith_2014}. Radio emission from stellar explosions is normally associated with an interaction between the fastest SN ejecta (i.e.\ the blastwave) and the wind-shaped CSM. 
As the forward shock propagates through the CSM, the electrons are accelerated, creating a bell-shaped non-thermal synchrotron spectrum \citep{Chevalier2017}.
The radio spectrum is characterized by a peak frequency, $\nu_{pk}$, that cascades to lower frequencies as the blastwave expands (e.g., \citealt{1998ApJ...499..810C,2006ApJ...651..381C}).
For synchrotron self-absorption (SSA) dominated spectra, \citet{1998ApJ...499..810C} suggests $\nu_{pk}\propto R^{2/7} B^{9/7}$, and the spectral peak flux $F_{pk}\propto R^{19/7}B^{19/7}$, where $R$ is the forward shock radius and $B$ is the post-shock magnetic field. 
The optically thin flux density at $\nu>\nu_{pk}$ scales as $F_{\nu,thin}{\propto}\,\nu^{-(p-1)/2}$ (where $p$ is the power-law index of the electron distribution, $N_{e}(\gamma_e){\propto}\,\gamma_e^{-p}$, and $\gamma_e$ is the electron Lorentz factor) and the optically thick spectrum at $\nu<\nu_{pk}$ is described as $F_{\nu,thick}\propto \nu^{5/2}$. 

During the SN interaction phase with a ``wind density profile'' environment expected around massive stars ($\rho_{\rm{CSM}}{\propto} r^{-2}$), the self-similar solutions by \cite{Chevalier82} apply and the shock radius evolves with time as $R\propto t^{\frac{n-3}{n-2}}$, where the density in the outer layers of the stellar progenitor has been parametrized as $\rho_{\rm SN}\propto r^{-n}$. 
In the limit of no evolution of the shock microphysical parameters (e.g., the fraction of post-shock energy in magnetic fields and relativistic electrons, $\epsilon_B$ and $\epsilon_e$), and using $n\approx 10$ as appropriate for compact massive stars \citep{2006ApJ...651..381C,1999ApJ...510..379M}, the equations above imply $\nu_{pk}\approx t^{-1}$ and $F_{pk}\approx$ constant. 
Since radio SNe typically show $p\approx 3$ (or $F_{\nu,thin}\propto \nu^{-1}$), $L_{\nu,pk}\le 10^{28}\,\rm{erg\,s^{-1}Hz^{-1}}$ and $\nu_{pk}\le 10$\, GHz at ${<}0.1$\,year after explosion, the prediction of this single wind model is $v_{pk}{\ll}1$\,GHz and a luminosity $<10^{27}\,\rm{erg\,s^{-1}Hz^{-1}}$ in the VLASS bandpass  at the current epoch (which corresponds to ${>}10^3$ days since explosion, Figure \ref{fig:lum_function}). 
We conclude that our sample of VLASS SNe require a deviation from a single-wind model. 

In the remainder of this section, we discuss three alternative explanations: (i) interaction of the SN shock with dense shells of CSM (Section \ref{subsec:overdensity}); (ii) emission from an off-axis relativistic jet entering our line of sight (Section \ref{subsec:jets}); (iii) emergence of emission from a PWN (Section \ref{subsec:PWN}).

\subsection{Dense detached CSM shells in the local SN environment}
\label{subsec:overdensity}
VLASS SNe show a level of radio emission comparable to the most luminous type IIn SNe (Fig. \ref{fig:lum_function}).
We place the VLASS-SNe into the phase space of radio observables $\nu_{pk}$, $L_{\nu,pk}$ and peak time $t_{pk}$ in Figure \ref{Fig:masslossSSA}, where  we calculated lines of constant shock velocity $v_{sh}$ and mass-loss $\dot M$ rate following the standard formulation of SSA radio emission from a blast wave during the interaction phase with  a wind-like environment \citep[e.g.,][]{1998ApJ...499..810C,2006ApJ...651..381C,2005ApJ...621..908S,Soderberg12}. 
Equipartition of energy between the relativistic electrons, protons and magnetic field, i.e.\ $\epsilon_e=\epsilon_B=1/3$, where $\epsilon_e$ is the fraction of thermal energy stored in electrons, and $\epsilon_B$ is the fraction of magnetic energy relative to the thermal energy leads to a solid lower limit on the mass-loss rate parameter $\dot M$ for a given wind velocity ($v_{w}$), where $\rho_{CSM}= \frac{\dot M}{4\pi v_{w}r^2}$. We present our results for both $\epsilon_e=\epsilon_B=1/3$ and for $\epsilon_e=0.1$ and $\epsilon_B=0.01$
Our discussion below focuses on our fiducial case of $\epsilon_e=0.1$ and $\epsilon_B=0.01$.
All $\dot M$ values quoted are for a wind velocity $v_w=10^3\,\rm{km\,s^{-1}}$.
A few considerations follow from Figure \ref{Fig:masslossSSA}:

\begin{figure*}[t!]
\vskip +0.0 true cm
\centering
  \includegraphics[width=1.05\columnwidth]{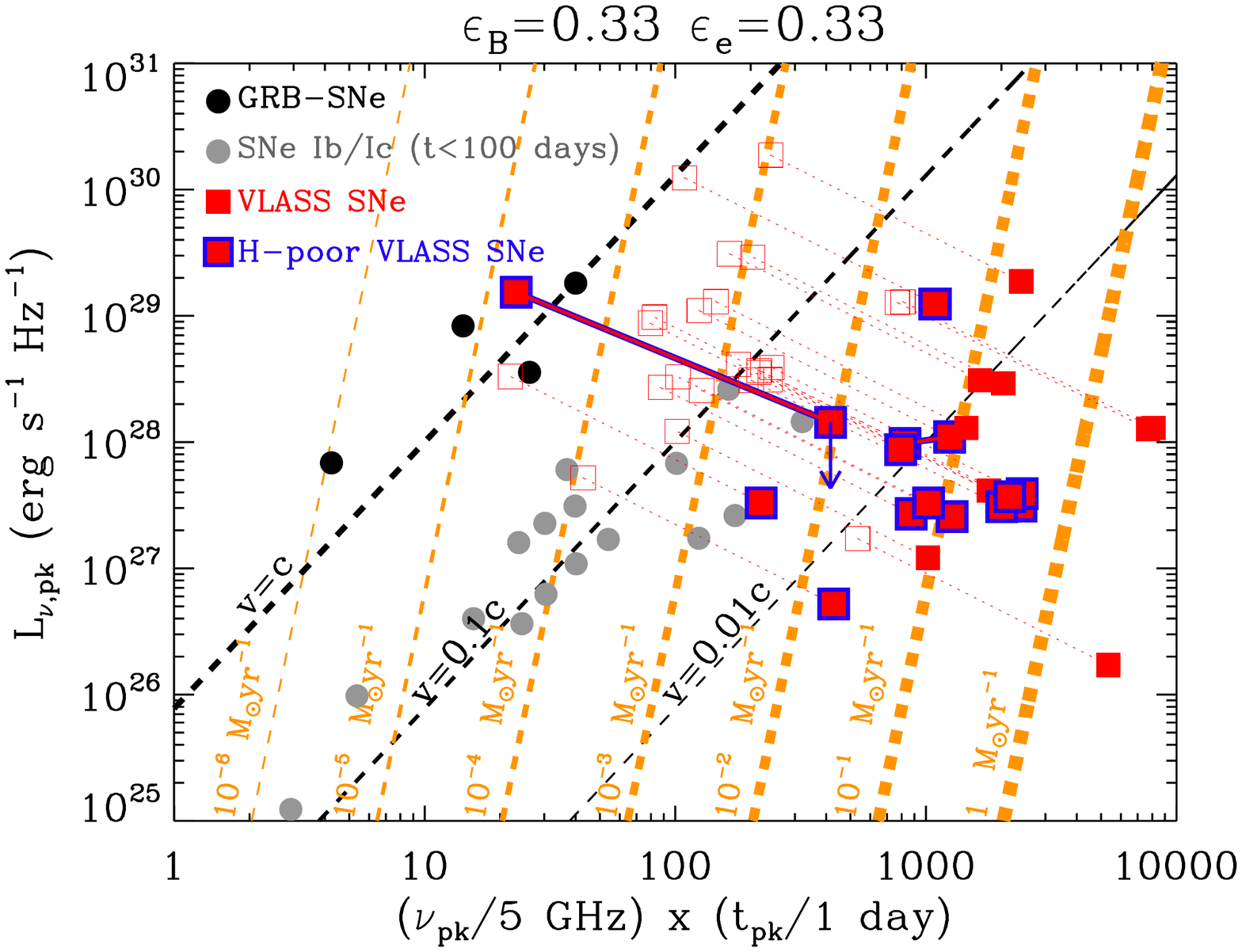}
    \includegraphics[width=1.05\columnwidth]{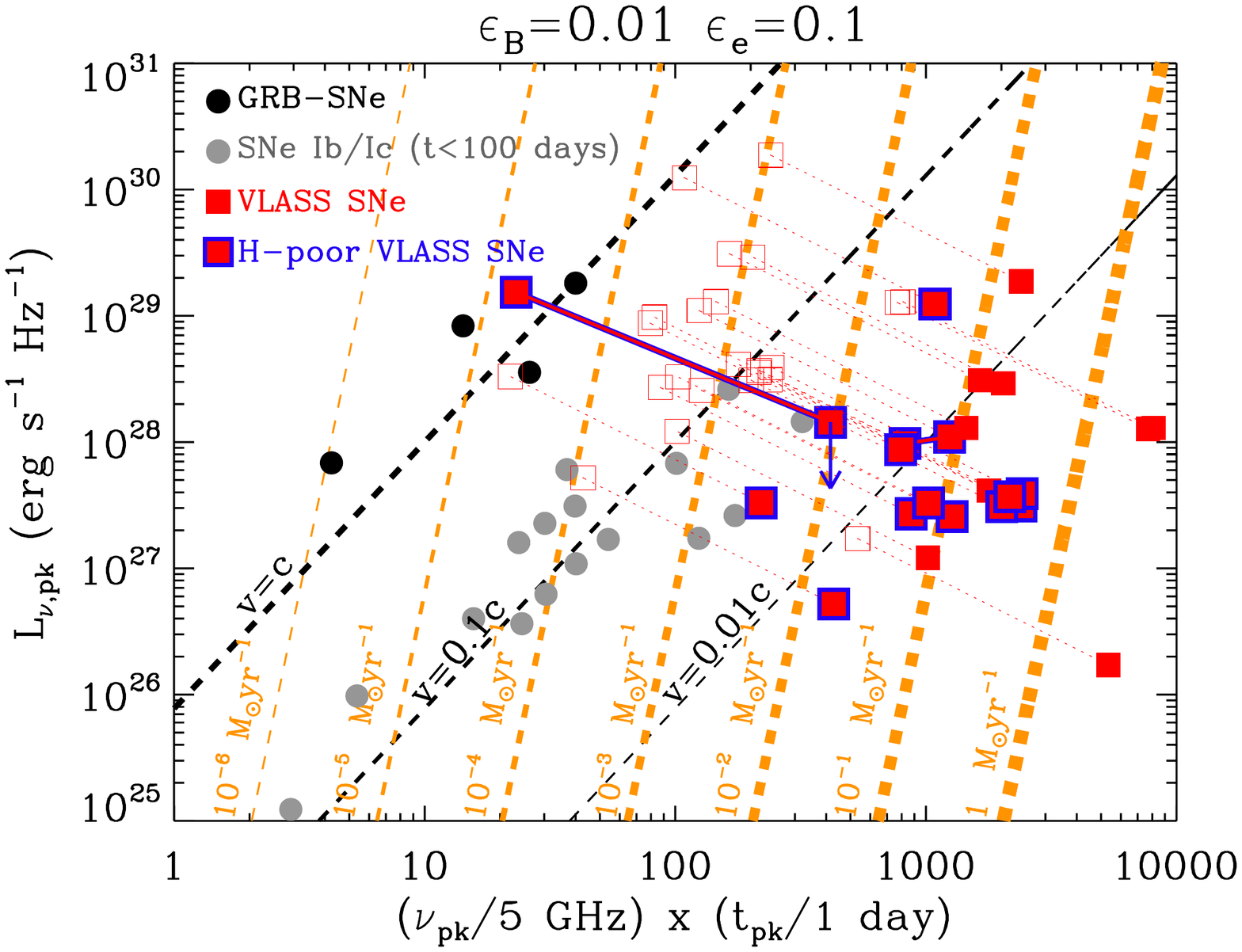}
  \vspace{-80pt}
  \caption{
  Location of the VLASS SNe (filled red squares)  in the phase space of radio observables. A blue outline marks the VLASS SNe with an H-poor spectrum at the time of explosion. 
  Black filled circles: GRB-SNe. Grey filled circles: H-stripped SNe from radio observations typically  acquired at $\lesssim 100$ days since explosion. 
  We assume $p=3$ and the shock microphysics indicated in the title of each plot. 
  Black dashed lines: lines of constant shock velocity assuming SSA only. 
  Orange dashed lines: lines of constant mass-loss rate, here calculated for an assumed wind velocity of $10^3$\,km/s. 
  The red filled squares show what the properties of the VLASS SNe would be in the case that the emission peaked in the VLASS band at the time of the observations (see Section \ref{subsec:overdensity}).
  Red open squares: location of the VLASS SNe for an optically thin spectrum $L_{\nu}\propto \nu^{-1}$, assuming that the $\nu_{pk}$ of the SSA  spectrum is below the VLASS band at $\approx 0.3$\,GHz. The VLASS object that crosses the $v=c$ line is SN\,2017iuk/GRB171205A.
  VLASS Memo 13 report an ${\sim}10\%$ overestimate in flux densities from the VLASS epoch 1 QuickLook data.
  No appreciable difference is found in this analysis when applying a 10\% correction to these figures.}
  References: \citet{Soderberg12} and references therein.
\label{Fig:masslossSSA}
\vspace{-0pt}
\end{figure*}

\begin{itemize}
    \item If $\nu_{pk}\gtrsim \nu_{obs}$ (where $\nu_{obs}$ is the frequency of the VLASS observations), then VLASS-SNe require very dense environments with an effective $\dot M\gtrsim 0.1\,\rm{M_{\odot}year^{-1}}$, which is significantly larger than the typical $\dot M$ inferred for non-type-IIn SN progenitors that comprise the majority of our sample (\citealt{smith_2014}). In absolute terms, the inferred $\dot M$ would compete with the most extreme mass-loss rates invoked for evolved massive stars.
    
    \item A lower $\nu_{pk}<\nu_{obs}$ would bring the VLASS-SNe in line with the lower $\dot M\sim 10^{-4}-10^{-3}\,\rm{M_{\odot}year^{-1}}$ that are typical of massive stars. The empty squares of Figure \ref{Fig:masslossSSA} show the location of VLASS-SNe for $\nu_{pk}=0.3$\,GHz as an example. However, the lower $\nu_{pk}$ would also lead to shock velocities $v_{sh}$ $\ge 0.1$c and $L_{\nu,pk}>10^{28}\,\rm{erg\,s^{-1}Hz^{-1}}$, implying that VLASS-SNe would constitute a class of radio SNe as luminous as long GRBs and with mildly relativistic shocks at ${>}10^3$\,days (and likely faster at earlier times).  
\end{itemize}

Since earlier-time radio follow up of some VLASS-SNe indicated ``normal'' SN shock speeds of $\sim 0.1-0.2c$ at a few months post-explosion (e.g., SN\,2012au in \citealt{Kamble2014}), it is clear that the relativistic ejecta scenario cannot explain the entire VLASS-SNe sample unless the relativistic ejecta is highly collimated (i.e.\ a jet) and pointing away from our line of sight at early times (i.e.\ off-axis). 
We further explore the relativistic ejecta scenario in Section \ref{subsec:jets}.

Mass-loss rates  $\dot M{\gtrsim} 0.1\,\rm{M_{\odot}year^{-1}}$, \emph{if} sustained until the time of explosion, would lead to very prominent type-IIn like spectroscopic features at earlier times for all the VLASS-SNe, which were not observed for the majority of the sample. 
Earlier radio observations of some targets also pointed to significantly lower $\dot M\approx 10^{-5}\,\rm{M_{\odot}year^{-1}}$  \citep[e.g., SNe 2004dk, 2012au and 2012ap;][]{2012ApJ...752...17W,Kamble2014,2015ApJ...805..187C} at the smaller radii probed at those epochs $r{\lesssim} 5\times 10^{16}\,\rm{cm}$.
The emerging picture is that at least some VLASS-SNe exploded in a low-density bubble surrounded by a shell of dense material at $r{\sim} v_{sh}\delta t = (v_{sh}/10^4\,\rm{km\,s^{-1}})(\delta t/8000\,\rm{days}){\approx}0.5$\,pc, consistent with the findings from the multi-wavelength monitoring of SNe 2003bg, 2004C, 2004dk, 2014C, and PTF11qcj (\citealt{Soderberg03bg}, \citealt{Margutti17}, \citealt{pooley2004dk}, \citealt{2014ApJ...782...42C}, \citealt{Palliyaguru2019}, \citealt{Murase2019}, \citealt{Brethauer2020}, \citealt{Balasubramanian_2021}, DeMarchi in prep.).
For VLASS-SNe from H-poor stellar progenitors (which interestingly dominate the sample), these overdensities might represent the shedding of their H-rich envelope in the centuries before core-collapse. 
Optical spectroscopy at the time of the radio re-brightenings of SN\,2003bg, SN\,2004dk, SN\,2014C and PTF11qcj confirmed the later appearance of H features in the spectra \citep{Soderberg03bg,pooley2004dk,Milisavljevic2015,Palliyaguru2019}, consistent with this scenario. 

Potential theoretical explanations of this phenomenology include the interaction of faster Wolf-Rayet winds with the slower winds of the red supergiant phase coupled with a shorter-than expected Wolf-Rayet phase; envelope ejection due to binary interaction; or mass shedding due to gravity-wave powered mass loss (e.g., \citealt{smith_2014}, \citealt{ZhaoFuller2020}, \citealt{WuFuller2021}).

\subsection{Off-axis relativistic jets}
\label{subsec:jets}

\begin{figure}[t!]
\centering
  \includegraphics[width=\columnwidth]{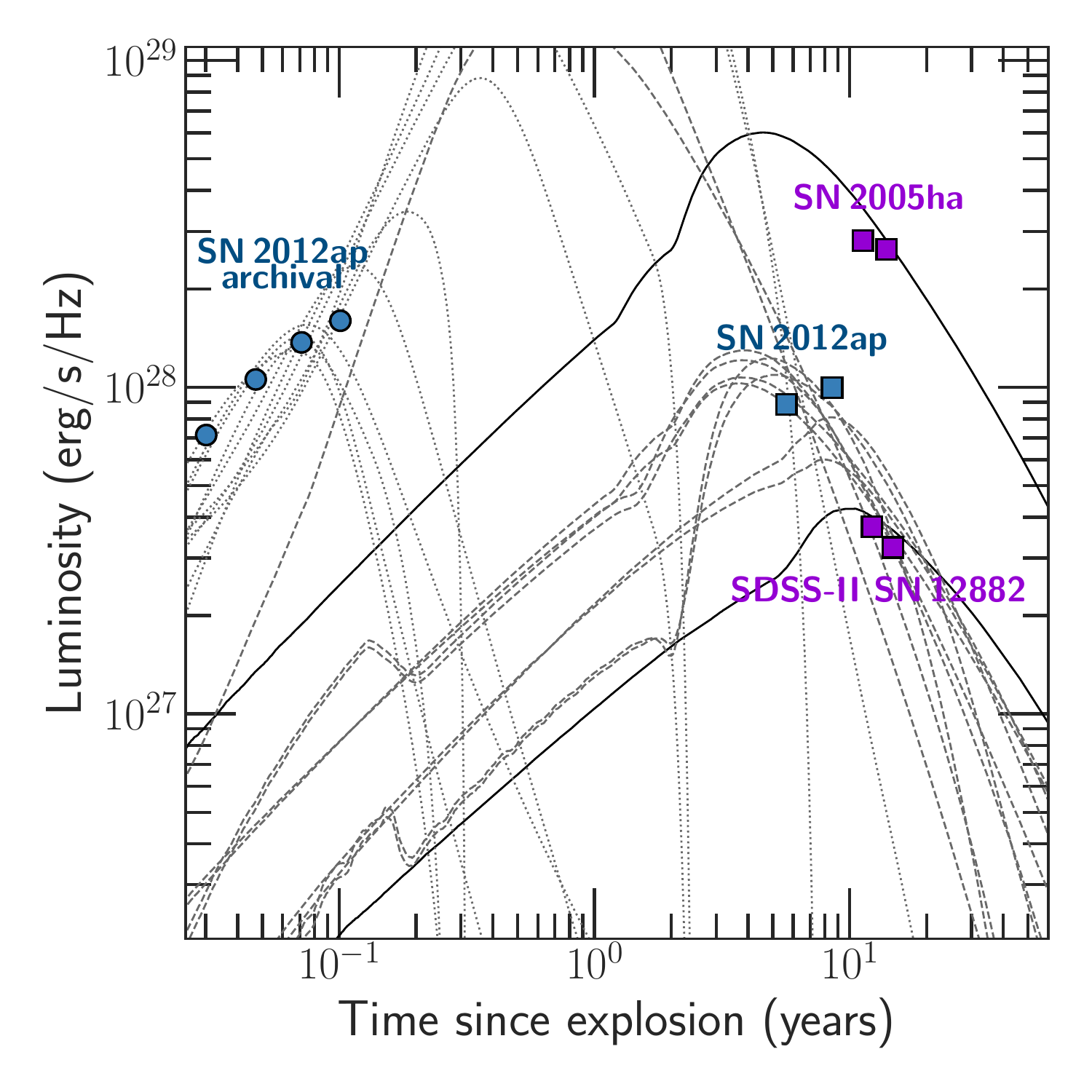}
  \caption{
  VLASS-SNe 3\,GHz light curves for SN\,2005ha, SN\,2012ap, and SDSS-II SN\,12882.
  The 3\,GHz SN\,2012ap archival light curve is included using the model from \citet{2015ApJ...805..187C}.
  The SN\,2012ap light curve is shown along with the 10 top-hat jet models in our grid that best fit the VLASS only light curve (gray dashed lines) and the 10 models that best fit the combined VLASS and archival light curve (gray dotted lines).
  SN\,2005ha and SDSS-II SN\,12882 are the only multi-epoch detected VLASS-SNe that are consistent with off-axis jet models from our grid (i.e. $\chi^2<1$).
  The solid black lines represent the off-axis jet models that are consistent with SN\,2005ha and SDSS-II SN\,12882 light curves.
  SN\,2005ha and SDSS-II SN\,12882 had limited spectroscopic follow-up and were not classified as H-rich or H-poor (e.g., \citealt{2005CBET..260....4M}).
  We show this figure as a proof of concept, but we note that the models that best fit the SN\,2005ha and SDSS-II SN\,12882 light curves have high isotropic kinetic energies of $E_{iso}=10^{55}$\,erg, corresponding to beaming-corrected energies of $E=3\times10^{53}$ and $1\times10^{54}$\,erg for SN\,2005ha and SDSS-II SN\,12882, respectively, and are likely unrealistic. 
  }
\label{fig:model_light_curves}
\end{figure}

Off-axis jets can result in bright synchrotron emission that peaks years after explosion \citep[e.g.,][]{Granot02,granot_2018}.
The emission from off-axis jets enters our line of sight as the jet decelerates in the ambient medium and relativistic beaming becomes less severe \citep{Rhoads97,Sari99}. 
The fraction of stellar explosions that are jet driven is still unclear (e.g., \citealt{Corsi_and_Lazatti_2021,2016ApJ...830...42C,Bietenholz_2014,2006ApJ...638..930S}).
Successful relativistic jet have been so far associated with broad-line type-Ic SNe accompanying cosmological GRBs, while partially successful and partially failed jets have been proposed to be powering low-luminosity GRBs and relativistic SNe, respectively (e.g., for observations see \citealt{Margutti_2014}, and for theory see \citealt{Morsony_2007,Lazzati_2012}). 
While observations of energetic H-stripped SNe point at a continuum of jet properties from normal Ibc SNe to GRB/SNe \citep[e.g.,][]{Xu_2008,Mazzali_2008,Margutti_2014c,Corsi_and_Lazatti_2021}, no bona fide off-axis jet has ever been associated with a SN without a GRB detection. 
In this context it is particularly interesting to note that the relativistic SN\,2012ap, which is one of only two known relativistic SNe without a GRB counterpart \citep{2015ApJ...805..187C,Soderberg_2010}, is detected by VLASS observations 5.7 and 8.5\,years after explosion.
Thus SN\,2012ap is a clear candidate for an off-axis jet driving late-time emission.

To determine whether the detected VLASS emission is associated with off-axis jets, we generated a set of synthetic 3\,GHz jet afterglow light-curves with Boxfit v2 \citep{2012ApJ...749...44V}. 
Boxfit assumes a top-hat jet structure, i.e.\ a jet with energy uniformly distributed within $\theta\le \theta_{jet}$ and $E=0$ for $\theta > \theta_{jet}$.
We assumed a wind-like CSM density profile. We explored the parameter space with a grid of parameter values defined as follows: isotropic-equivalent jet kinetic energies of $E_{iso} = [10^{50},10^{51},10^{52},10^{53},10^{54},10^{55}]\,\rm{erg}$; jet opening angles of $\theta_{jet} =[ 5^\circ,15^\circ,30^\circ]$; off-axis angle $\theta_{obs} =[30^\circ,60^\circ,90^\circ]$ from our line of sight; mass-loss rates of $\dot M=[10^{-8},10^{-7},10^{-6},10^{-5},10^{-4} 10^{-3}]$\,$M_{\odot}$\,year$^{-1}$ for $v_w=1000\,\rm{km\,s^{-1}}$; shock micro-physical  parameters $\epsilon_{e} = 0.1$, $\epsilon_{B} = [0.001, 0.01]$, $p=[2,2.5,3]$. Finally we used the $\chi^2$ as a metrics to evaluate the agreement between the models and the VLASS data of Figure \ref{fig:lum_function}.

None of the VLASS-SNe with archival (i.e.\ pre-VLASS) data points are consistent with our model off-axis jet light curves. 
We find that the synthetic models that best approximate the VLASS data of SDSS-II SN\,12882, SN\,2002hi, SN\,2005ha, SN\,2009fi, and SN\,2012cc with $\chi^2\lesssim 1$ have large off-axis angles $\theta_{obs}\ge 60^\circ$, large densities corresponding to $\dot M>10^{-5}$\,$M_{\odot}$\,year$^{-1}$, $p\sim 3$, $\epsilon_B=0.01$ coupled with large $E_{iso}\ge 10^{54}\,\rm{erg}$ and large jet angle $\theta_{jet}\ge 15^\circ$. The values of these parameters is driven by the large radio luminosities of the sample at late times, and imply extremely large beaming-corrected jet energies $3\times 10^{52}-10^{54}\,\rm{erg}$. While we show some examples of top-hat off-axis jet light-curve consistent with the VLASS data in Figure \ref{fig:model_light_curves}, we consider this top-hat jet scenario unlikely because of the large jet energies needed and the fact that only SNe with a sparse data set can be fitted. We consider alternative jet models and environments below.

We start by noting that in SN\,1965G, SN\,2004C, SN\,2005ha, SN\,2012ap, SN\,2012at, and SDSS-II SN\,12882, the radio flux density remains nearly constant over ${\sim}2-3$\,years between two VLASS epochs. 
SN\,1986J also has nearly constant radio flux densities (see Figure \ref{fig:lum_function}); however, it has only been observed in a single VLASS epoch.
Numerical simulations of GRB jets propagating through a stratified media show that nearly flat, wide peaks are obtained only if the jet propagates through a wind-profile medium with $\rho\propto r^{-2}$ \citep[see Figure 1 in][]{granot_2018}. Jets propagating through a uniform density environment have a much more narrow peak (e.g., as seen in the GRB\,170817A afterglow \citealt{Margutti_2020arXiv}) and are ruled out by our observations. GRB\,170817A also clearly showed that relativistic jets can have angular structure (i.e.\ the jet is not necessarily top-hat; see e.g., \citealt{2020PhR...886....1N} and references therein).

The propagation of relativistic GRB jets through a massive  Wolf-Rayet progenitor star leads to the production of extended wide-angle outflows known as cocoons, with masses ${\approx}10^{-2}-10^{-1}$\,M$_{\odot}$ and energies ${\approx} 10^{50}-10^{51}$\,ergs \citep[see, e.g.,][]{Lazzati_2005,nakar_2017,DeColle_2021} possibly observationally identified in SN\,2017iuk/GRB171205A \citep{Izzo_2019}.
Once the jet breaks out of the star, the cocoon engulfs the star and expands nearly spherically into the environment (see e.g.\ Figure 3 of \citealt{DeColle_2021}). The cocoon initially expands with relativistic velocities (corresponding to Lorentz factors ${\sim}2-10$), but later decelerates to mildly relativistic velocities at ${\sim}10^{16}$\,cm \citep[see Figure 2 in][]{decolle_2018}. 
Particle acceleration through the shock cocoon itself will lead to a bright afterglow. 
While GRB jets are collimated and enter into the observer line of sight only at late times, the cocoon radio emission should be detectable at early times by observers located at nearly all angles (as beaming effects are much less important in the slower moving cocoon material) and would thus be able to explain the larger radio fluxes of pre-VLASS observations. 
The predicted early-time radio emission \citep{decolle_2018,DeColle_2021} is similar to that observed in relativistic SNe 2009bb and 2012ap \citep{Soderberg_2010,Margutti_2014,2015ApJ...805..187C}.
Several SNe in our sample, including SN\,2012ap, have been observed at early times (${\sim}$days to a month after explosion), but only SN\,2012ap showed mildly relativistic material consistent with the expectations from the cocoon model.
SN\,2012ap is the only VLASS-SN for which a cocoon and off-axis relativistic jet is a viable explanation.
The largely uncollimated, mildly relativistic cocoon would be responsible for the early emission.
The late-time VLASS emission would be powered by the off-axis relativistic jet.
In this case, SN\,2012ap would represent the first evidence of a cocoon and jet system from a massive stellar explosion.
Future multifrequency observations will test this scenario.
Thus, with the exception of SN\,2012ap, we find that the late-time VLASS emission is unlikely to be caused by relativistic jets.

\subsection{Emergence of emission from a pulsar wind nebula}
\label{subsec:PWN}
Another candidate for the cause of late-time radio emission from SNe is the presence of a PWN (e.g., \citealt{Gaensler_2006, Slane2017}).  Core-collapse SNe, which comprise the totality of our sample, are expected to leave a compact remnant. 
If a fast rotating NS is left behind, it can feed a steady highly-energetic wind of relativistic particles into the SN ejecta, and this ``bubble'' of relativistic particles is referred to as a PWN. 
As this wind interacts with the slower SN ejecta, a termination shock forms and high-energy photon emission heats and ionizes the surrounding SN ejecta. 
Shortly after the explosion, the emission is absorbed by the dense ejecta \citep[e.g.,][]{Metzger_2014,Murase_2015,Murase_2016,Murase_2021}.
Over time, as the ejecta expands and the optical depth decreases, the PWN emission becomes observable. 
No SN has unambiguously shown the transition from ejecta-dominated emission to PWN-dominated emission.
Recently, there have been hints towards the detection of a PWN associated with SN\,1987A.
This suggestion is due to non-thermal emission in the hard X-rays \citep{Greco2021}, and from the radio detection of a warm dust concentration at the center of the remnant \citep{Cigan2019}; however, alternative mechanisms to explain the emission cannot be ruled out.
Beyond SN\,1987A, two young SNe have been suggested to harbor PWNe (SN\,1986J and SN\,2012au), and both are in our VLASS-SNe sample.

The presence of a PWN energizing the ejecta in a young SN has been proposed to explain the anomalous state of high ionization inferred from optical spectroscopy of the H-stripped energetic SN\,2012au ${\approx}6$\,years after explosion by \cite{Milisavljevic18}.
The spectra of this transient acquired $\approx$7 years after explosion were dominated by forbidden oxygen lines with velocities of  ${\approx}2300$\,km\,s$^{-1}$. 
Oxygen resides in the inner part of the SN ejecta, thus one explanation for this emission is the presence of a pulsar that ionizes the internal material \citep{Milisavljevic18}.
The lack of narrow hydrogen in the early spectra of SN\,2012au suggests a different powering mechanism than CSM-ejecta interaction, and supports the scenario of ionization by a pulsar as the origin of the emission. 

\citet{Bietenholz2002} suggested the late-time radio emission from SN\,1986J is evidence of a PWN.
SN\,1986J showed a broad radio SED $7-16$ years after explosion with a spectrum at $\nu>10$\,GHz which evolved from thin to thick (i.e.\ an inverted radio spectrum). 
However, observed SEDs of evolved PWNe are relatively flat, with typical spectral indices between $-0.3$ and $0.0$.
In contrast, SN\,1986J has an SED that peaked at ${\approx}20$\,GHz, with an absorbed optically thick region, and an optically thin spectral index of $-0.76$.
SN\,2012au has a similarly shaped SED at $8$\,years post-explosion (Terreran et al.\ in prep.).
The bell-like synchrotron SEDs produced by CSM-interaction of the SN shock wave peak below GHz frequencies on these time-scales. 
The observed radio spectrum of SN\,1986J and SN\,2012au is also unusual for evolved PWNe, but we emphasize that the spectral properties of nascent PWNe that are a few years old are not observationally well constrained. 
From a theoretical perspective, we expect the radiative electrons to be in the ``fast-cooling'' regime, which can lead to radio spectra similar to those observed \citep[e.g.,][]{Murase_2016,Omand_2018,Murase_2021}.
A young PWN is expected to be smaller in size than the SN ejecta, thus one can distinguish between the shock interaction and a PWN with very long baseline interferometry.

Interestingly, both SN\,1986J and SN\,2012au are in the VLASS-SNe sample.
If the bright radio emission is confirmed to be powered by PWNe, the associated PWNe would be the two youngest discovered to date.
No forming PWN has been observed in the Milky Way or the Magellanic Clouds.
The youngest known galactic PWN, Kes 75, has an estimated age of $480\pm40$\,years \citep{Reynolds2018}, and thus little is known about PWN properties in the years to decades after the SN explosion.

\section{Conclusions}
\label{sec:conclusions}
We present evidence for a population of 19 radio luminous  SNe ($L_{\nu}\sim10^{26}-10^{29}$\,erg\,s$^{-1}$\,Hz$^{-1}$ at 3\,GHz) ${\approx}1-60$\,years after explosion found in  the first epoch of the VLASS. This is part of the radio phase space of stellar explosions that has not been systematically explored so far. 
Our filtering procedure leveraged multiwavelength catalogs to remove potential AGN contaminants, and other known radio sources leading to a sample that is entirely comprised of core-collapse SNe and surprisingly dominated by stellar explosions with hydrogen stripped progenitors at the time of collapse.  Our main result is that the large radio luminosities at these late stages of evolution require deviation from the traditional single wind mass-loss scenario and/or spherical shock assumption. Potential alternatives include the following:
\begin{enumerate}
    \item Initial expansion of the SN shock into a lower-density bubble, followed by strong shock interaction with a sharp density increase (i.e. a ``bubble'' plus  detached shell CSM structure). This dense shell might be connected to the shedding of the H-rich stellar envelope in the centuries before core-collapse through mass-loss mechanisms that have yet to be observationally identified. VLASS SNe are as luminous as the most luminous radio SNe IIn few yrs post explosion, which indicates CSM densities at large radii from the progenitors that are comparable to those inferred for SNe IIn.

    \item While top-hat relativistic jets viewed off-axis are unlikely to provide an adequate explanation due to the under-prediction of the pre-VLASS radio observations of most elements of the sample, relativistic jets with structure are not ruled out. SN\,2012ap, which showed evidence for an uncollimated mildly relativistic outflow at $\delta t< 40$ days, is the primary candidate for being the first jet+cocoon system in a massive star observed off-axis, which may signal that relativistic SNe are ``cocoons'' observed early on.
    \item The final alternative is the emergence of a PWN. The VLASS-SNe sample includes SN\,1986J and SN\,2012au, the two young SNe that have previously been suggested to have PWNe powered late-time radio emission.
\end{enumerate}

The VLA Sky Survey provides an unprecedented and unbiased window into the variable radio sky, combining the large survey area of the Northern VLA Sky Survey with the depth and angular resolution of the Faint Images of the Radio Sky survey. These features, and the planned multiple field visits, are particularly useful for the discovery and study of extragalaxtic transients, where the angular resolution (and higher frequency, 3\,GHz vs.\ 1\,GHz) minimizes confusion by the host galaxies of transients of interest. Planned interferometers such as the Next Generation VLA \citep{2015arXiv151006438C} and the Square Kilometer Array \citep{2009IEEEP..97.1482D} will expand our ability to study the variable radio sky with increased depth. The VLASS is complimented by other surveys and serendipitous transient discovery programs being carried out with SKA pathfinder instruments such as ASKAP (VAST; \citealt{murphy2013}), Westerbork (Apertif; \citealt{adams2019}), MeerKAT (ThunderKAT; \citealt{fender}), LOFAR (LoTSS; \citealt{shimwell2017}), which encompass a range of frequencies and angular resolutions while providing access to the southern sky.  

Follow-up with multiwavelength observations including radio spectral energy distributions, and optical spectroscopy will help constrain the mechanisms responsible for the bright radio emission of our VLASS-SN sample.
We will present the multiwavelength follow-up of the VLASS detected SNe sample in future papers.


\software{
APLpy \citep{aplpy2012,aplpy2019},
Astropy \citep{astropy:2013, astropy:2018},
BOXFIT \citep{2012ApJ...749...44V},
Pandas \citep{mckinney-proc-scipy-2010},
PyBDSF \citep{2015ascl.soft02007M},
Q3C \citep{2006ASPC..351..735K},
SExtractor \citep{1996A&AS..117..393B}
}

\acknowledgements
We thank the referee for providing constructive comments.
We also thank Seth Bruzewski for providing astrometric corrections to the VLASS Quicklook epoch 1 data.

This work is supported by the Heising-Simons Foundation under grant \#2018-0911 (PI: Margutti). R.M.\ acknowledges support by the NSF under grants AST-1909796 and AST-1944985. 
F.D.C.\ acknowledges support from the UNAM-PAPIIT grant AG100820.
R.B.D.\ acknowledges support from National Science Foundation (NSF) under grant 1816694.
The National Radio Astronomy Observatory is a facility of the National Science Foundation operated under cooperative agreement by Associated Universities, Inc.
This publication makes use of data products from the Wide-field Infrared Survey Explorer, which is a joint project of the University of California, Los Angeles, and the Jet Propulsion Laboratory/California Institute of Technology, and NEOWISE, which is a project of the Jet Propulsion Laboratory/California Institute of Technology. WISE and NEOWISE are funded by the National Aeronautics and Space Administration.
This research has made use of data obtained from the Chandra Source Catalog, provided by the Chandra X-ray Center (CXC) as part of the Chandra Data Archive.
This research has made use of data obtained from the 4XMM XMM-Newton serendipitous source catalogue compiled by the 10 institutes of the XMM-Newton Survey Science Centre selected by ESA.
This work made use of data supplied by the UK Swift Science Data Centre at the University of Leicester.
This research was supported in part through the computational resources and staff contributions provided for the Quest high performance computing facility at Northwestern University which is jointly supported by the Office of the Provost, the Office for Research, and Northwestern University Information Technology.
Development of the BOXFIT code was supported in part by NASA through grant NNX10AF62G issued through the Astrophysics Theory Program and by the NSF through grant AST-1009863.
This research made use of APLpy, an open-source plotting package for Python \citep{aplpy2012,aplpy2019}.


\appendix
\section{VLASS detected SNe}
We present in Table \ref{tab:final_detection_list} the sample of SNe detected in the VLASS data set.
The second column lists whether the progenitor was hydrogen rich or hydrogen poor at the time of explosion.
Flux densities and positions were derived as described in Section \ref{sec:sample_final_list}, with any deviations from the default procedure given in the `Notes' column. 
The list of VLASS SNe is ordered by increasing right ascension. 
\citet{2021ApJ...914...42B} calculated the astrometric corrections required to align the VLASS Quicklook epoch 1 data with the Gaia catalog, and we list the coordinates that include these astrometric corrections.
The positional errors include the uncertainties from PyBDSF fits and the astrometric corrections (Bruzewski private communication) added in quadrature.
We applied the \citet{2021ApJ...914...42B} derived corrections and astrometric uncertainties for the VLASS epoch 2.1 observations.
We note that applying the VLASS epoch 1 uncertainties may overstate the positional uncertainties in the second epoch observations, since the second VLASS epoch will have likely benefited from studying the systematic uncertainties in the first epoch (see e.g.\ VLASS Memo 13\footnote{The VLASS Project Memo Series is listed at https://go.nrao.edu/vlass-memos.}).

We report the \texttt{PyBDSF} flux density errors which are purely statistical.
There are known flux density offsets in the VLASS Quicklook images, as detailed in VLASS Memo 13 and the CIRADA Catalogue User Guide.
The detection type defines the nature of the source structure, where ``S'' indicates a single Gaussian that is the only source in the island, ``C'' indicates a single source in an island with other sources, and ``M'' indicates multiple Gaussian source.
The angular separation lists the distance between the listed VLASS position and the optical position.
For the SN classifications, ``Pec'' and ``BL'' stand for peculiar and broad-lined, respectively.
Four SNe (20\% of the sample) had limited follow-up leaving the classification unknown, but they are believed to be core-collapse SNe (i.e.\ SN\,1965G, SN\,2005ha, SDSS-II SN\,8524, and SDSS-II SN\,12882).

\begin{longrotatetable}
\begin{deluxetable*}{llllrrrrrr}

\tablecaption{
Supernovae detected in VLASS epoch 1
\label{tab:final_detection_list}
}

\tablehead{
\nocolhead{Name} & \nocolhead{Progenitor} & \multicolumn5c{VLASS} & \nocolhead{Luminosity} & \colhead{Angular} & \nocolhead{Classification} \\
\cline{3-7}
\colhead{Name} & \colhead{Progenitor} & \colhead{R.A.} & \colhead{Decl.} & \colhead{Flux Density} & \colhead{Detection} & \colhead{Obs.\ Date} & \colhead{Luminosity} & \colhead{Separation} & \colhead{Classification} \\
\nocolhead{Name} & \colhead{[H-rich/poor]} & \colhead{[hh:mm:ss.ss]} & \colhead{[dd:mm:ss.ss]} & \colhead{[mJy]} & \colhead{[S, M or C]} & \colhead{[MJD]} & \colhead{[erg/s/Hz]} & \colhead{[$''$]} & \nocolhead{Classification} }

\startdata
        SN\,1986J & ? & 02:22:31.293(15) & $+$42:19:57.5(3) & $1.3\pm0.2$ $\phantom{^{x}}$ & C & 58588 & $(1.6\pm0.2)\times10^{26}$ & $0.56$ & IIn$\phantom{^{x}}$ \\
        SN\,2017hcb & H-poor & 02:36:23.756(14) & $+$31:42:36.2(3) & $0.5\pm0.2$ $\phantom{^{x}}$ & M & 58569 & $(3.0\pm1.2)\times10^{27}$ & $0.99$ & Ib$\phantom{^{x}}$ \\
        SDSS-II SN\,12882 & ? & 03:03:49.977(9) & $-$00:12:14.3(3) & $1.7\pm0.4$ $^{a}$ & S & 58103 & $(2.8\pm0.7)\times10^{28}$ & $0.32$ & ?$\phantom{^{x}}$ \\
         & & 03:03:49.975(10) & $-$00:12:14.2(3) & $1.6\pm0.3$ $^{a}$ & S & 59078 & $(2.6\pm0.5)\times10^{28}$ & $0.41$ & \\
        SN\,2003bg & H-poor & 04:10:59.436(6) & $-$31:24:50.2(3) & $4.1\pm0.2$ $\phantom{^{x}}$ & S & 58663 & $(2.83\pm0.14)\times10^{27}$ & $0.61$ & IcPecBL$\phantom{^{x}}$ \\
        SN\,2012at & H-poor & 04:54:52.783(7) & $-$37:19:16.9(3) & $2.0\pm0.2$ $\phantom{^{x}}$ & S & 58153 & $(2.5\pm0.2)\times10^{27}$ & $0.41$ & Ic$\phantom{^{x}}$ \\
         & & 04:54:52.786(8) & $-$37:19:17.4(3) & $1.9\pm0.2$ $\phantom{^{x}}$ & S & 59155 & $(2.3\pm0.2)\times10^{27}$ & $0.70$ & \\
        SN\,2012ap & H-poor & 05:00:13.734(5) & $-$03:20:51.4(3) & $4.0\pm0.3$ $\phantom{^{x}}$ & S & 58027 & $(8.9\pm0.7)\times10^{27}$ & $0.25$ & IcBL$\phantom{^{x}}$ \\
         & & 05:00:13.738(5) & $-$03:20:51.6(3) & $4.5\pm0.3$ $\phantom{^{x}}$ & S & 59078 & $(1.00\pm0.07)\times10^{28}$ & $0.23$ & \\
        SN\,2005ha & ? & 06:21:49.110(6) & $+$00:21:56.2(3) & $2.2\pm0.2$ $\phantom{^{x}}$ & S & 58123 & $(3.7\pm0.3)\times10^{27}$ & $1.23$ & ?$\phantom{^{x}}$ \\
         & & 06:21:49.106(6) & $+$00:21:56.0(3) & $1.9\pm0.3$ $\phantom{^{x}}$ & S & 59048 & $(3.2\pm0.5)\times10^{27}$ & $1.09$ & \\
        SN\,2002hi & H-rich & 07:19:54.127(9) & $+$17:58:18.5(3) & $2.1\pm0.3$ $\phantom{^{x}}$ & S & 58572 & $(1.7\pm0.2)\times10^{29}$ & $0.72$ & IIn$\phantom{^{x}}$ \\
        SN\,2017iuk & H-poor & 11:09:39.519(5) & $-$12:35:18.5(3) & $4.8\pm0.2$ $\phantom{^{x}}$ & S & 58150 & $(1.39\pm0.06)\times10^{29}$ & $0.24$ & IcBL$\phantom{^{x}}$ \\
         & & \nodata & \nodata & $<0.45$ $\phantom{^{x}}$ & \nodata & 59133 & $<1.3\times10^{28}$ & \nodata & \\
        SN\,2004C & H-poor & 11:27:29.80(2) & $+$56:52:47.9(3) & $4.2\pm0.6$ $^{b}$ & S & 58020 & $(2.8\pm0.4)\times10^{27}$ & $0.57$ & Ic$\phantom{^{x}}$ \\
         & & 11:27:29.77(2) & $+$56:52:47.9(3) & $5.3\pm0.9$ $\phantom{^{x}}$ & S & 59064 & $(3.5\pm0.6)\times10^{27}$ & $0.79$ & \\
        SN\,1965G & ? & 12:11:54.049(5) & $+$24:06:58.5(3) & $7.7\pm0.4$ $\phantom{^{x}}$ & S & 58082 & $(1.15\pm0.06)\times10^{28}$ & $2.59$ & ?$\phantom{^{x}}$ \\
         & & 12:11:54.045(5) & $+$24:06:58.4(3) & $7.8\pm0.3$ $\phantom{^{x}}$ & S & 59099 & $(1.17\pm0.05)\times10^{28}$ & $2.52$ & \\
        SN\,2012cc & H-rich & 12:26:56.829(9) & $+$15:02:45.6(3) & $2.3\pm0.4$ $\phantom{^{x}}$ & S & 58590 & $(1.10\pm0.19)\times10^{27}$ & $0.36$ & II$\phantom{^{x}}$ \\
        SN\,2012au & H-poor & 12:54:52.257(5) & $-$10:14:50.5(3) & $4.5\pm0.3$ $\phantom{^{x}}$ & S & 58553 & $(3.0\pm0.2)\times10^{27}$ & $1.16$ & Ib$\phantom{^{x}}$ \\
        PTF11qcj & H-poor & 13:13:41.480(9) & $+$47:17:56.8(3) & $6.8\pm0.2$ $\phantom{^{x}}$ & S & 58561 & $(1.12\pm0.03)\times10^{29}$ & $0.44$ & IcBL$\phantom{^{x}}$ \\
        SN\,2009fi & H-rich & 14:06:05.757(6) & $+$11:47:13.6(3) & $2.2\pm0.2$ $\phantom{^{x}}$ & S & 58611 & $(1.17\pm0.11)\times10^{28}$ & $0.88$ & IIb$\phantom{^{x}}$ \\
        SN\,2004dk & H-poor & 16:21:48.872(4) & $-$02:16:17.6(3) & $6.3\pm0.2$ $\phantom{^{x}}$ & S & 58624 & $(3.34\pm0.11)\times10^{27}$ & $0.75$ & Ib$\phantom{^{x}}$ \\
        SDSS-II SN\,8524 & ? & 21:29:23.354(6) & $+$00:56:42.9(3) & $1.7\pm0.2$ $\phantom{^{x}}$ & S & 58023 & \nodata$^{c}$ & $1.18$ & ?$\phantom{^{x}}$ \\
         & & \nodata & \nodata & $<0.5$ $\phantom{^{x}}$ & \nodata & 59049 & \nodata$^{c}$ & \nodata & \\
        SN\,2016coi & H-poor & 21:59:04.127(8) & $+$18:11:10.8(3) & $1.8\pm0.2$ $\phantom{^{x}}$ & S & 58604 & $(4.7\pm0.5)\times10^{26}$ & $0.35$ & IcBL$\phantom{^{x}}$ \\
        SN\,2014C & H-poor & 22:37:05.601(6) & $+$34:24:31.5(3) & $29.0\pm0.3$ $\phantom{^{x}}$ & S & 58642 & $(7.91\pm0.08)\times10^{27}$ & $0.49$ & Ib$^{d}$ \\
\enddata
\tablecomments{
a) The region surrounding SDSS-II SN\,12882 is contaminated by radial artifacts from quasar PB\,6989, so flux density could be less reliable. \\
b) The region surrounding SN\,2004C is contaminated by radial artifacts from NVSS J112731$+$565240, so the flux density could be less reliable. SN\,2004C is clearly part of an extended emission complex, which is not detected with \texttt{PyBDSF} unless the island and detection threshold are both lowered. \\
There is a clear point source at the location of SN\,2004C which we associate with the transient, but the flux density here is likely unreliable.\\
{c) SDSS-II SN\,8524 has no known redshift, thus a luminosity cannot be calculated.}\\
d) SN\,2014C was initially classified as a type Ib but was later classified as a type IIn.}
\end{deluxetable*}
\end{longrotatetable}

\bibliography{mybib}{}
\bibliographystyle{aasjournal}

\end{document}